\documentclass[10pt, a4paper]{article}
\pdfoutput=1
\usepackage[utf8]{inputenc}
\usepackage[T1]{fontenc}
\usepackage{lmodern, textcomp}
\usepackage[american]{babel}
\usepackage{csquotes}

\usepackage{eurosym}
\usepackage[nottoc]{tocbibind}
\usepackage{authblk}
\usepackage{fancyhdr}
\usepackage{setspace}
\usepackage{titlesec}
\usepackage{changepage}
\usepackage{lastpage}
\usepackage{xspace}

\usepackage{physics}
\usepackage{graphicx}
\usepackage{subcaption}
\usepackage{float}
\usepackage{makecell}
\usepackage{booktabs}
\usepackage{multirow}
\usepackage{marginnote}
\usepackage{enumitem}
\usepackage[multiple]{footmisc}

\usepackage[usenames, dvipsnames, svgnames, table]{xcolor}

\usepackage[backend=bibtex8,
            citestyle=numeric-comp,
            maxbibnames=99,
			sorting=none,
			sortcase=false,
			sortcites=true,
			giveninits=true
			]{biblatex}

\usepackage{amsmath, amsfonts, amssymb}
\usepackage{mathtools}
\usepackage{cancel}
\usepackage[
	print-unity-mantissa=false, separate-uncertainty=true
]{siunitx}
\usepackage{listings}

\usepackage[pdftex,
            breaklinks=true,
            linktocpage,
            colorlinks=true,
            urlcolor=blue,
            linkcolor=blue,
            citecolor=red
            ]{hyperref}

\newcommand{\email}[1]{\href{mailto:#1}{\nolinkurl{#1}}}
\newcommand{\emailfoot}[1]{\thanks{\email{#1}}}

\newcounter{draftcommentcnt}
\NewDocumentCommand{\draftcomment}{s O{red} m}{%
	\def\margnote{\IfBooleanTF{#1}{\marginnote}{\marginpar}}%
	\stepcounter{draftcommentcnt}%
	\textcolor{#2}{#3}%
	\margnote{\textcolor{#2}{$\Leftarrow$ \arabic{draftcommentcnt}}}%
}

\numberwithin{equation}{section}

\usepackage[sort&compress, english]{cleveref}

\usepackage[heightrounded, margin=0.75in,footskip=0.25in]{geometry}
\fancypagestyle{preprint}{
	\fancyhf{}

}

\input{arxiv.def}

\hypersetup{pdftitle={Deep learning complete intersection Calabi-Yau manifolds}}

\title{\bf Deep learning complete intersection Calabi-Yau manifolds}

\author[1,2,3]{Harold Erbin\emailfoot{erbin@mit.edu}}
\author[3,4]{Riccardo Finotello\emailfoot{riccardo.finotello@cea.fr}}

\affil[1]{%
	Center for Theoretical Physics, Massachusetts Institute of Technology
	\protect\\
	Cambridge, MA 02139, \textsc{Usa}
}

\affil[2]{%
	\textsc{Nsf Ai} Institute for Artificial Intelligence and Fundamental Interactions
}

\affil[3]{%
	Université Paris Saclay, \textsc{Cea}, \textsc{List}
	\protect\\
	Palaiseau, F-91120, France
}

\affil[4]{%
	Université Paris Saclay, \textsc{Cea}, Service d'Études Analytiques et de Réactivité des Surfaces (\textsc{Sears})
	\protect\\
	Gif-sur-Yvette, F-91191, France
}

\newcommand{\SU}[1]{\ensuremath{\mathrm{SU}(#1)}\xspace}
\newcommand{\h}[2]{\ensuremath{h^{( #1, #2 )}}\xspace}

\begin{document}

\maketitle
\begin{abstract}
We review advancements in deep learning techniques for complete intersection Calabi-Yau (CICY) 3- and 4-folds, with the aim of understanding better how to handle algebraic topological data with machine learning. We first discuss methodological aspects and data analysis, before describing neural networks architectures. Then, we describe the state-of-the art accuracy in predicting Hodge numbers. We include new results on extrapolating predictions from low to high Hodge numbers, and conversely.
\end{abstract}
\vspace{0.5cm}

\thispagestyle{preprint}

\hrule
\pdfbookmark[1]{\contentsname}{toc}
\tableofcontents
\bigskip
\hrule
\bigskip

\section{Introduction}

In recent years, deep learning has become a relevant research theme in physics and mathematics.
It is a very efficient method for data processing, and elaboration and exploration of patterns~\cite{goodfellow2016deep}.
Though the basic building blocks are not new~\cite{rosenblatt1958perceptron}, the increase in computational capabilities and the creation of larger databases lead new deep learning techniques to thrive.
Specifically, the understanding of the geometrical structures~\cite{bronstein2017geometric, Bronstein:2021:GeometricDeepLearning} and the representation learning~\cite{bengio2013representation} are of particular interest from a mathematical and theoretical physics points of view~\cite{ruehle2020data, He:2020mgx, He:2021oav, He:2021:CalabiYauLandscapeGeometry}.

We are interested in applications of data science and deep learning techniques for algebraic topology, and especially Hodge numbers, of complete intersection Calabi-Yau (CICY) manifolds~\cite{green1987calabi, candelas1988complete, Green:1989:AllHodgeNumbers}.
Traditional methods from algebraic topology lead to complicated algorithms, without closed-form solutions in general.
Hence, it is interesting to derive new computational methods: given its track-record, deep learning is particularly promising as a path to the discovery of novel structures and analytic formulas, but also to the classification of Calabi-Yau manifolds~\cite{hubsch1992calabi, Anderson:2018:TASILecturesGeometric, he2020calabi}, which is still an important open mathematical problem.
However, a first step is to be able to reproduce known results, which is non-trivial given that algebraic topology provides a genuinely new type of data compared to what is usually studied by computer scientists.
The computation using deep learning has been studied in~\cite{He:2017:MachinelearningStringLandscape, Bull:2018:MachineLearningCICY, Bull:2019:GettingCICYHigh, He:2021:MachineLearningCalabiYau} (see~\cite{Ruehle:2017:EvolvingNeuralNetworks, He:2019:DistinguishingEllipticFibrations, Krippendorf:2020:DetectingSymmetriesNeural, Larfors:2020:ExploreExploitHeterotic, He:2021:WorldGrainSand} for other works related to CICY) before being almost completely solved by~\cite{erbin2021inception, erbin2021machine, erbin2021deep}: the objective of this paper is to review these state-of-the art results.

CICYs are also relevant for string theory model building where CY manifolds are needed to describe the compactified dimensions.
The general properties of the 4-dimensional effective field theories can be determined from the analysis of the topology~\cite{Grana:2006:FluxCompactificationsString}.
Given the complexity of string vacua~\cite{denef2007computational, halverson2019computational}, deep learning techniques may enable faster computations and may grant a larger exploration of possibilities.

CICY 3- and 4-folds have already been entirely classified and all of their topological properties are known:
\begin{itemize}
    \item in complex dimension 3, there are \num{7890} CICYs which have been represented in two different ways: the \emph{original} dataset~\cite{green1987calabi, candelas1988complete, Green:1989:AllHodgeNumbers} contains the original list of CICYs, while a newer classification~\cite{anderson2017fibrations} uses the \emph{favourable} representation whenever possible. We shall focus on the former, the first being the most difficult case from a machine learning point of view;
    \item in complex dimension 4, \num{921497} distinct CICYs were classified~\cite{gray2013all, gray2014topological}.
\end{itemize}
In this sense, they represent the ideal benchmark to test learning algorithms in supervised tasks.

\section{Data analysis of CICYs}

CY $N$-folds are $N$-dimensional complex manifolds with \SU{N} holonomy, or, equivalently, with a vanishing first Chern class~\cite{hubsch1992calabi}.
They are characterised by their topological properties, such as the Hodge numbers and the Euler characteristic.
These features directly translate into properties of the 4-dimensional effective action, such as the number of chiral multiplets in heterotic compactifications, and the number of hyper and vector multiplets in type II compactifications.
Ultimately, these are connected to the number of fermion generations, which could be used to test the effectiveness of the models.

The simplest CYs are constructed as \emph{complete intersections} of hypersurfaces in a product of complex projective spaces $\mathbb{P}^{n_1} \times \dots \times \mathbb{P}^{n_m}$~\cite{green1987calabi}.
They are defined by systems of homogeneous polynomial equations, whose solutions identify CY manifolds.
As we are interested in classifying CYs as topological manifolds, it is sufficient to keep track only of the dimensions of the projective spaces and the degree of the equations.
In the general case of $m$ projective spaces and $k$ equations, a CICY $X$ is represented by a \emph{configuration matrix} of integer entries:
\begin{equation}
    \label{eq:config-matrix}
    X =
    \begin{bmatrix}
    \begin{tabular}{c|ccc}
        $\mathbb{P}^{n_1}$ & $\alpha^1_1$ & $\cdots$ & $\alpha^1_k$ \\
        $\vdots$           & $\vdots$     & $\ddots$ & $\vdots$     \\
        $\mathbb{P}^{n_m}$ & $\alpha^m_1$ & $\cdots$ & $\alpha^m_k$ \\
    \end{tabular}
    \end{bmatrix},
\end{equation}
where $\alpha^i_r$ are positive integers satisfying:
\begin{equation}
    n_i + 1 = \sum\limits_{r = 1}^k \alpha^i_r,
\end{equation}
encoding the vanishing of the first Chern class, and
\begin{equation}
    \dim\limits_{\mathbb{C}} X = \sum\limits_{i = 1}^{m} n_i - k = N,
\end{equation}
where, in the following, $N = 3,\, 4$.
Notice that different configuration matrices can describe the same topological manifold.
First, any permutation of lines and columns does not modify the intersection of the hypersurfaces.
Second, different intersections can define the same manifold.
Such ambiguity is often fixed by imposing some ordering on the coefficients.
Moreover, in some cases and optimal representation of $X$ is available: in its \emph{favourable} representation, the configuration matrix enables easier computations of topological quantities.

The classification of CICY 3-folds has been dealt with in~\cite{candelas1988complete}, which lead to a dataset of \num{7890} configuration matrices, \num{22} of which in block diagonal form.
In this dataset, \SI{62}{\percent} of the configuration matrices are in a favourable representation, in which $\h{1}{1} = m$.
More recently, a different dataset of CICY 3-folds has been produced~\cite{anderson2017fibrations}: in this case, \SI{99}{\percent} of the manifolds are in favourable representation.
In what follows, we focus on the first, original dataset as it represents a more challenging and fascinating scenario from a machine learning point of view.
As such, we deal with configuration matrices whose maximal size is $12 \times 15$ (the most general representation uses $15 \times 18$ matrices).
CICY 4-folds have also been classified in~\cite{gray2013all, gray2014topological}, producing \num{921497} distinct matrices, with \num{15813} manifolds in block diagonal form, of maximal size $16 \times 20$.
Here, \SI{55}{\percent} of the matrices are in a favourable representation.
In our analysis, we discard the block diagonal matrices, as they can be understood from lower-dimensional manifolds.

\subsection{Hodge number distributions}

\begin{figure*}
    \centering
    \begin{subfigure}[b]{0.475\linewidth}
    \centering
    \includegraphics[width=\linewidth]{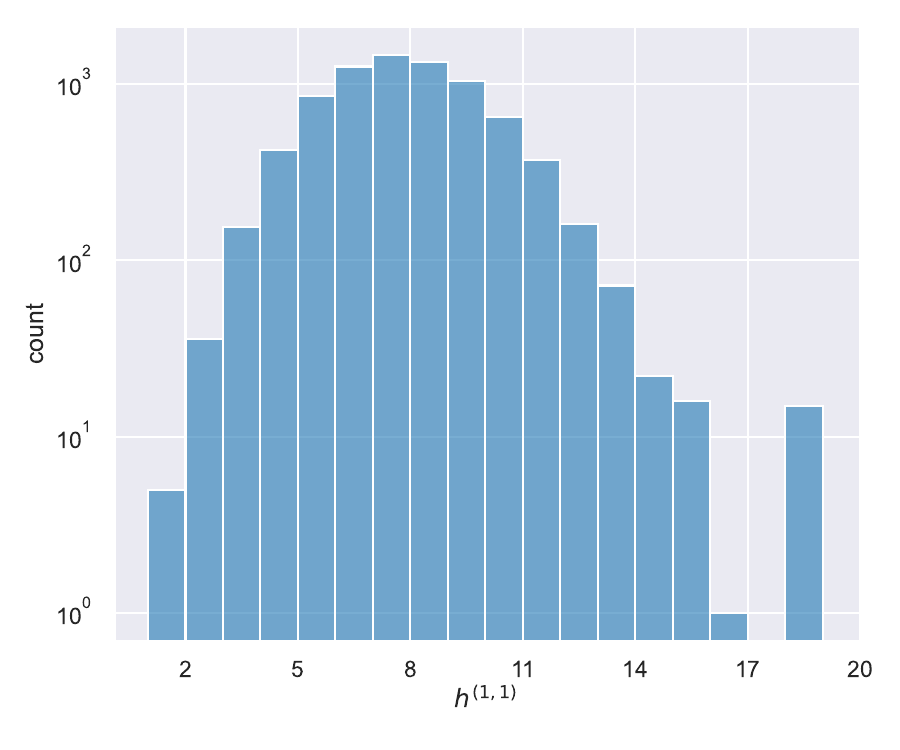}
    \caption{\h{1}{1}}
    \end{subfigure}
    \hfill
    \begin{subfigure}[b]{0.475\linewidth}
    \centering
    \includegraphics[width=\linewidth]{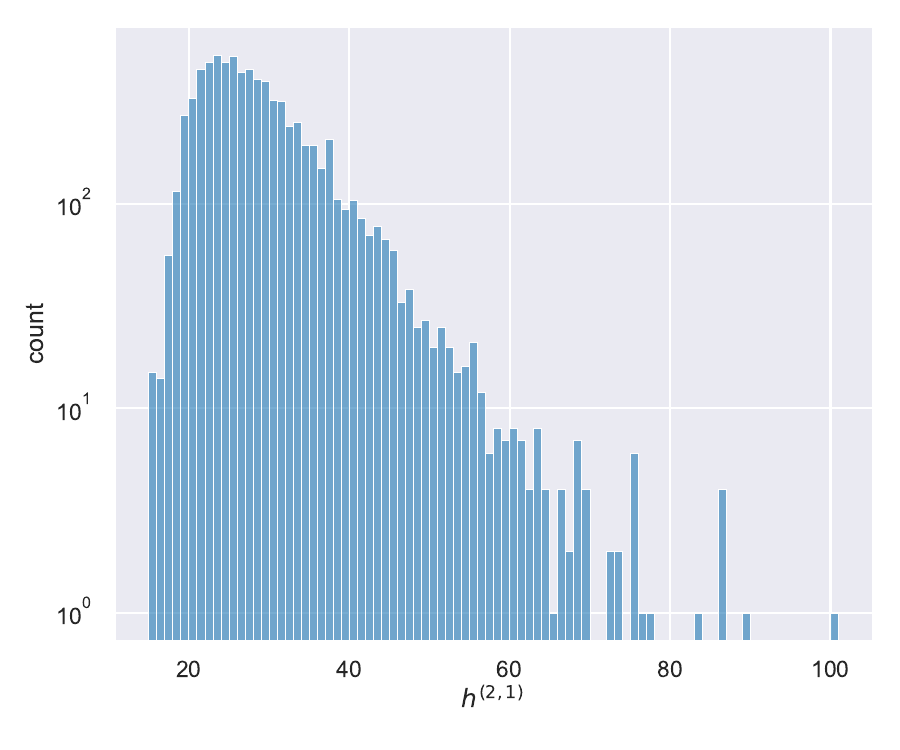}
    \caption{\h{2}{1}}
    \end{subfigure}
    \caption{Hodge numbers for CICY 3-folds.}
    \label{fig:cicy3_hodge}
\end{figure*}

The number of Hodge numbers depend on the complex dimension $N$ of $X$.
In the case most relevant to string theory, $\dim\limits_{\mathbb{C}} X = 3$ implies the existence of only two non-trivial Hodge numbers, \h{1}{1} and \h{2}{1}, whose distributions are shown in~\Cref{fig:cicy3_hodge}.
Their average, minimum and maximum values are:
\begin{equation}
    \expval{\h{1}{1}} = 7.4^{19}_{1},
    \quad
    \expval{\h{2}{1}} = 29^{101}_{15}.
\end{equation}
As visible in the figures, the two distributions greatly differ, with \h{1}{1} being almost normally distributed, contrary to \h{2}{1} which presents also several missing values towards the upper limit of its range.

\begin{figure*}
    \centering
    \begin{subfigure}[b]{0.475\linewidth}
    \centering
    \includegraphics[width=\linewidth]{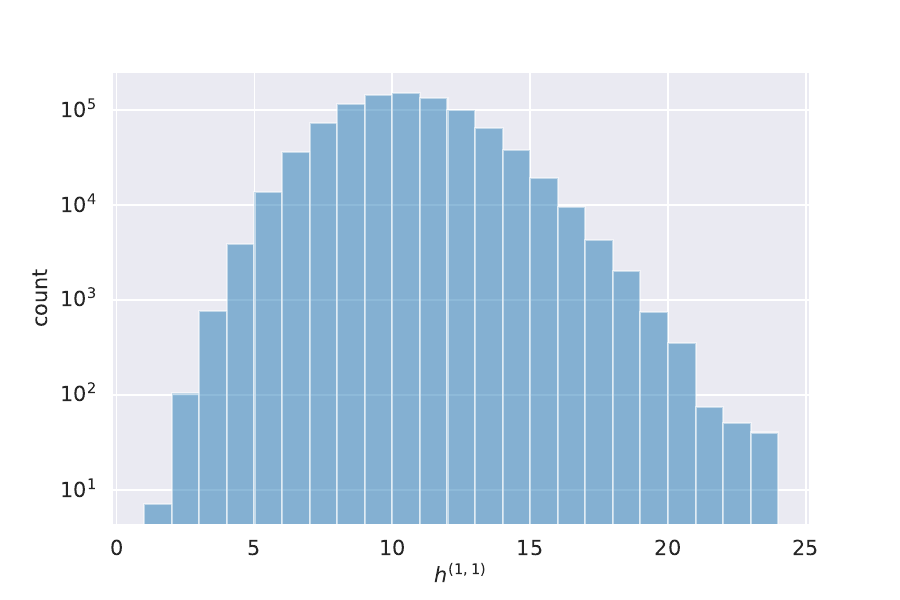}
    \caption{\h{1}{1}}
    \end{subfigure}
    \hfill
    \begin{subfigure}[b]{0.475\linewidth}
    \centering
    \includegraphics[width=\linewidth]{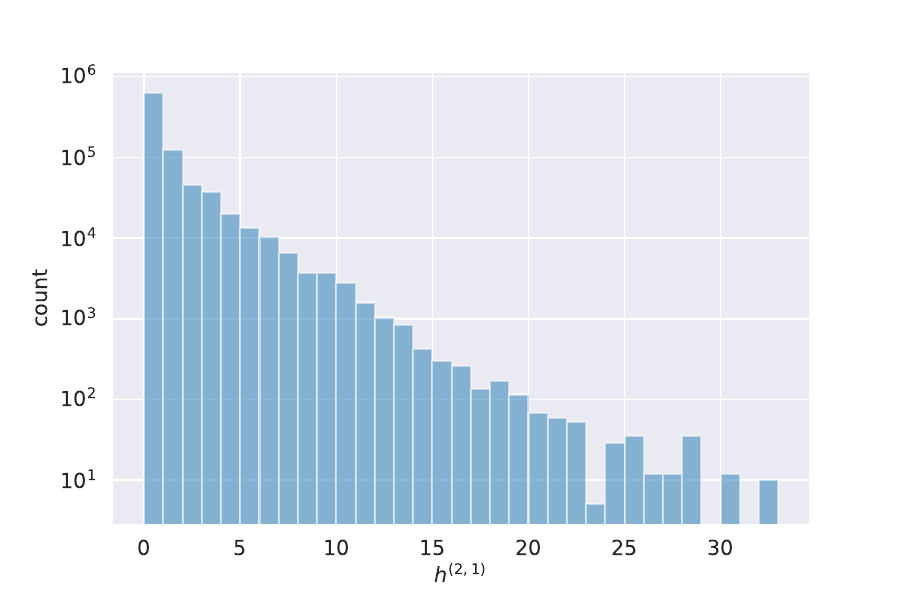}
    \caption{\h{2}{1}}
    \end{subfigure}
    \\
    \begin{subfigure}[b]{0.475\linewidth}
    \centering
    \includegraphics[width=\linewidth]{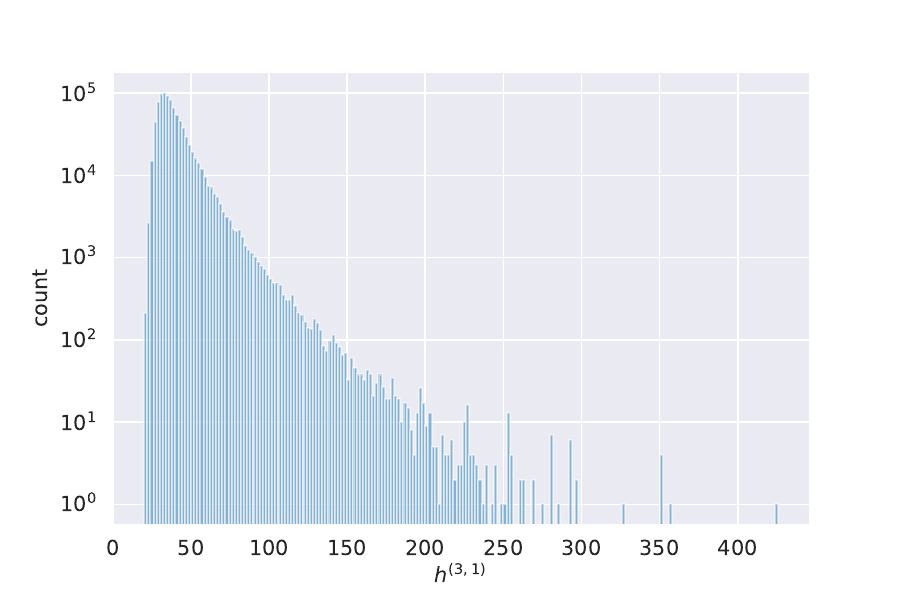}
    \caption{\h{3}{1}}
    \end{subfigure}
    \hfill
    \begin{subfigure}[b]{0.475\linewidth}
    \centering
    \includegraphics[width=\linewidth]{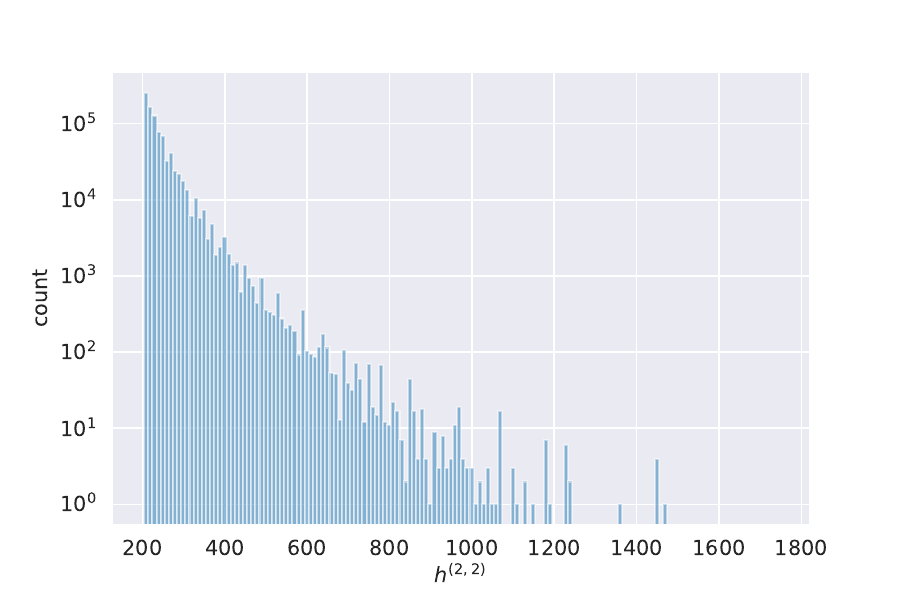}
    \caption{\h{2}{2}}
    \end{subfigure}
    \caption{Hodge numbers for CICY 4-folds.}
    \label{fig:cicy4_hodge}
\end{figure*}

CICY 4-folds are used for M- and F-theory compactifications: in this case, there are four non-trivial Hodge numbers whose distributions are shown in~\Cref{fig:cicy4_hodge}.
As for the previous case, their average values and ranges are:
\begin{equation}
\begin{split}
    \expval{\h{1}{1}} = 10^{24}_{1},
    & \quad
    \expval{\h{2}{1}} = 0.8^{33}_{0},
    \\
    \expval{\h{3}{1}} = 40^{426}_{20},
    & \quad
    \expval{\h{2}{2}} = 240^{1752}_{204}.
\end{split}
\end{equation}
Notice that, in this case, the distributions are highly unbalanced and defined on vastly different ranges.
For instance, \h{2}{1} vanishes for \SI{70}{\percent} of the configuration matrices in the dataset.
Moreover, \h{2}{1} and \h{2}{2} look more similar to exponential distributions, which, in turn, may complicate the learning process.
Note that the Hodge numbers are not independent~\cite{gray2014topological}:
\begin{equation}
    - 4 \h{1}{1} + 2 \h{2}{1} - 4 \h{3}{1} + \h{2}{2} = 44.
\end{equation}

In each case, a linear combination of the Hodge numbers provides another topological number, the Euler characteristics, which is much simpler.
For the 4-folds, we have
\begin{equation}
    \chi = 4 + 2 \h{1}{1} - 4 \h{2}{1} + 2 \h{3}{1} + \h{2}{2},
\end{equation}
while for the 3-folds the formula reads:
\begin{equation}
    \chi = 2 ( \h{1}{1} - \h{2}{1} ).
\end{equation}
Note that they do not provide constraints on the Hodge numbers strictly speaking, since the Euler number is not a fixed number but depends on the CY.

\subsection{Data engineering}

\begin{figure}[tp]
    \centering
    \includegraphics[width=0.5\linewidth]{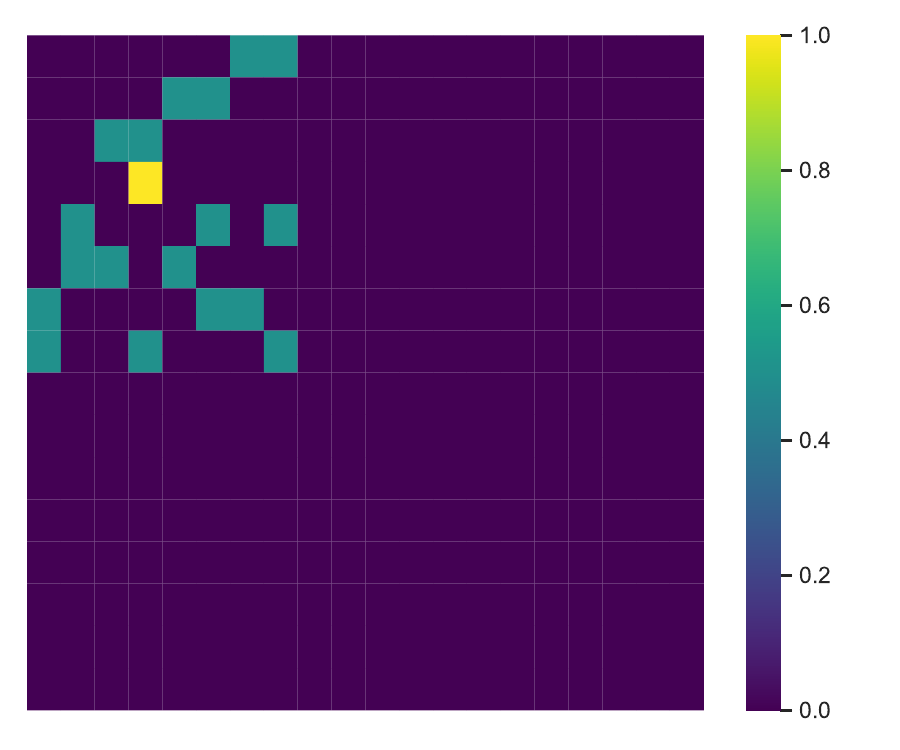}
    \caption{Configuration matrix of a CICY 4-folds ($\h{1}{1} = 8$, $\h{2}{1} = 0$, $\h{3}{1} = 38$, $\h{2}{2} = 228$) normalized by the highest possible entry in the training set.}
    \label{fig:cicy4_heatmap}
\end{figure}

In order to better understand patterns and characteristics of the data, first, we proceed to a phase of exploratory data analysis~\cite{erbin2021machine}.
In turn, the in-depth study of the CICY datasets helps in designing the learning algorithms.
As a reference, we adopt the machine learning dictionary and call \emph{features} the input variables, and \emph{labels} the target predictions, for simplicity.
Notice that the following analysis should be performed on the subset of configuration matrices used for training the learning algorithms: in this case, we are authorised to access the full information on features and labels, which should not be touched in the case of the test set.

\begin{figure}[tbp]
    \centering
    \includegraphics[width=0.7\linewidth, trim={6in 0 0 0.35in}, clip]{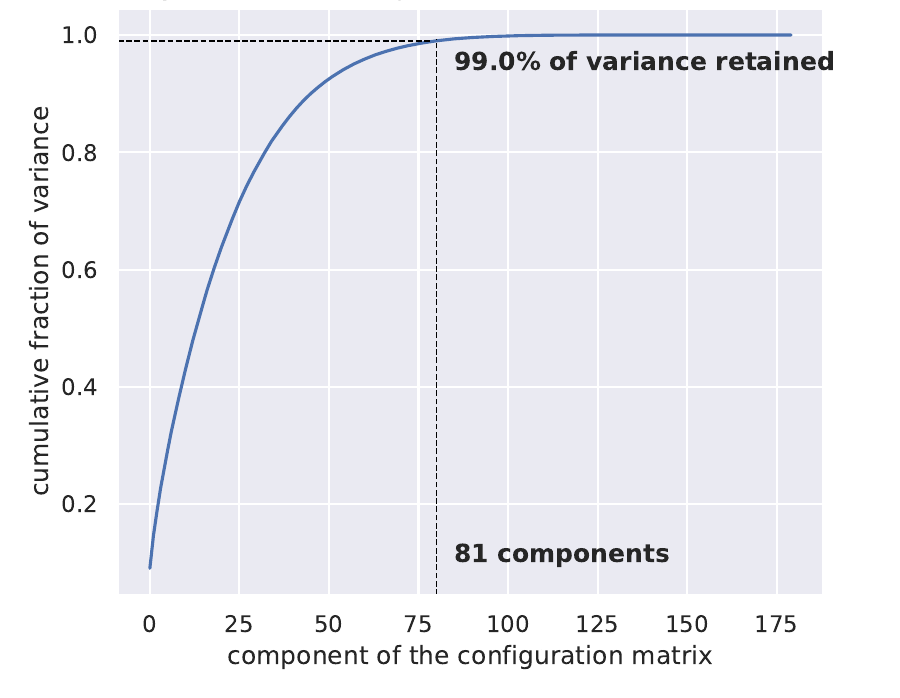}
    \caption{Singular value decomposition of CICY 3-folds.}
    \label{fig:cicy3_svd}
\end{figure}

We start from  a phase of \emph{feature engineering}~\cite{zheng2018feature}, in which new input features are derived from the raw input, that is the configuration matrix (see~\Cref{fig:cicy4_heatmap}).
Engineered features are redundant variables, encoding information already present in the input data, under a different representation.
They may help during the learning phase by providing an alternative formulation of salient characteristics.
In the case CICYs, useful quantities may be~\cite{hubsch1992calabi}:
\begin{itemize}
    \item the number of projective spaces $m$,
    \item the number of equations $k$,
    \item the number $f$ of $\mathbb{P}^1$,
    \item the number of $\mathbb{P}^2$,
    \item the number $F$ of $\mathbb{P}^n$, with $n \neq 1$,
    \item the excess number $N_{ex} = \sum\limits_{r = 1}^F \left( n_r + f + m - 2 k \right)$,
    \item the dimension of the cohomology group $\mathrm{H}^{(0)}$ of the ambient space,
    \item the Frobenius norm of the matrix,
    \item the list of dimensions of the projective spaces and statistics (mean, min, max, median),
    \item the list of degree of the equations and statistics (mean, min, max, median),
    \item K-means clustering of the matrices (with a variable number of clusters),
    \item principal components of the configuration matrix.
\end{itemize}
The Principal Components Analysis (PCA), through the study of the singular value decomposition,  may lead to clues for the learning algorithms.
As shown in~\Cref{fig:cicy3_svd} in the case of CICY 3-folds, the majority of the variance is contained in a smaller number of entries of the configuration matrices.
Ultimately, this may help in finding the optimal learning algorithms: the information is contained in smaller portions of the configuration matrix, which can thus be analysed in smaller patches.

\begin{figure}[tbp]
    \centering
    \includegraphics[width=0.7\linewidth, trim={0 0 0 0.35in}, clip]{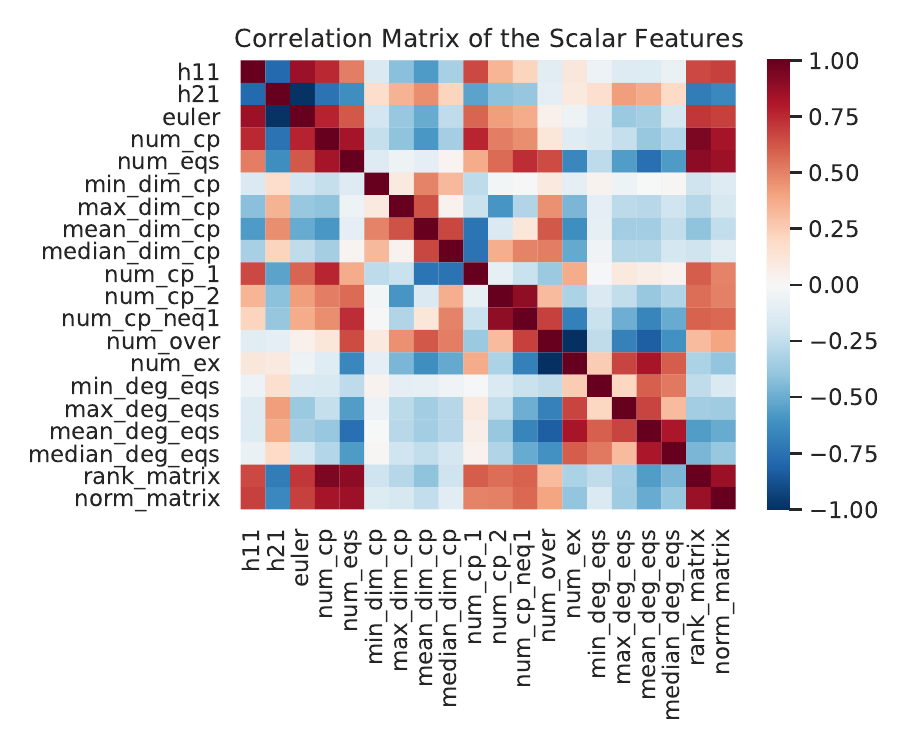}
    \caption{Correlation matrix of engineered features for CICY 3-folds.}
    \label{fig:cicy3_corr}
\end{figure}

The next important step is the \emph{feature selection}, that is, the extraction of a smaller number of features which may be more relevant to the determination of the labels.
To get an idea, the correlations between engineered features may give an indication on the linear dependence of the features.
\Cref{fig:cicy3_corr} shows the case of the CICY 3-folds: the number of projective spaces $m$, together with the rank and norm of the configuration matrix, seem to show a good correlation with the labels.

\begin{figure}[tbp]
    \centering
    \includegraphics[width=0.7\linewidth, trim={0 0 6in 0.35in}, clip]{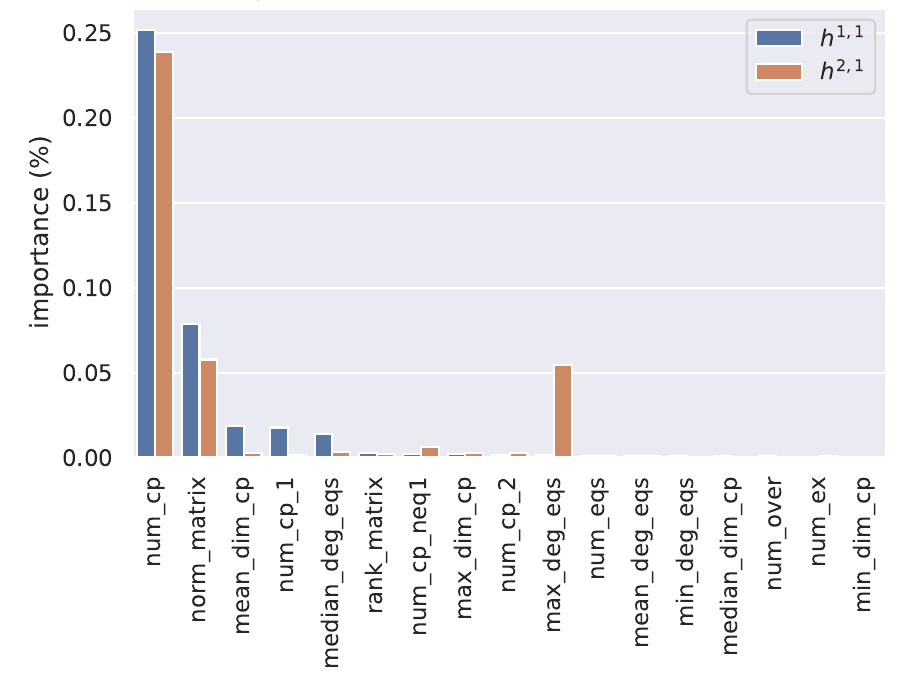}
    \caption{Importance of the scalar features for CICY 3-folds.}
    \label{fig:cicy3_scalar_imp}
\end{figure}

\begin{figure}[tbp]
    \centering
    \includegraphics[width=0.7\linewidth, trim={6in 0 0 0.35in}, clip]{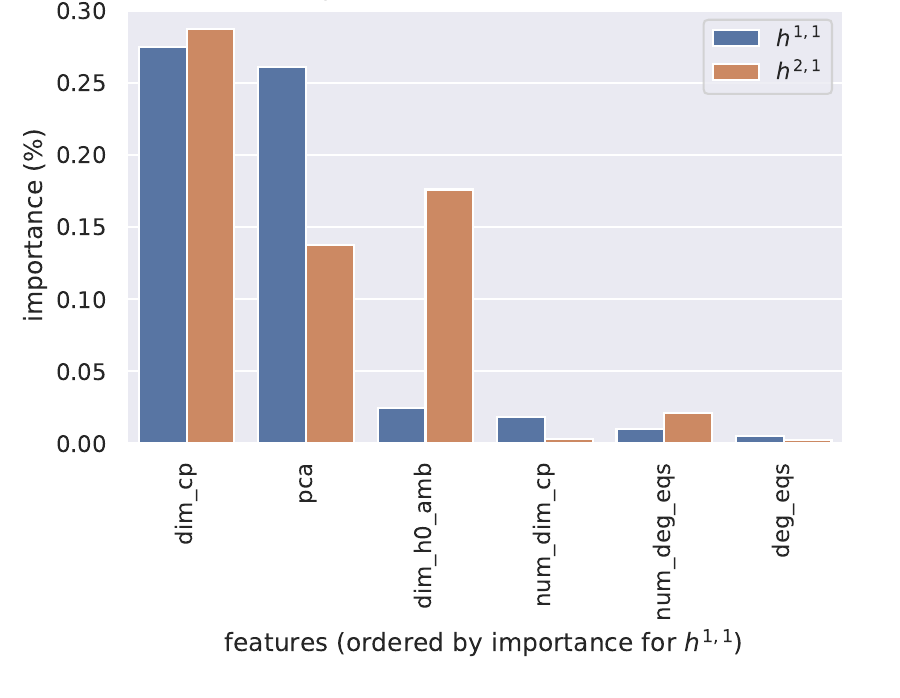}
    \caption{Importance of the vector features for CICY 3-folds.}
    \label{fig:cicy3_vector_imp}
\end{figure}

Other strategies for the selection of the features can also be used.
For instance, decision trees~\cite{steinberg2009cart} may return relevant information: as their binary structure is based on the optimal split of the variables to classify or predict the labels, the analysis of the choices made by the algorithm naturally selects the most important features.
In~\Cref{fig:cicy3_scalar_imp}, we show the importance of the scalar engineered features: the number of projective spaces effectively seems to be relevant for predicting the labels.
\Cref{fig:cicy3_vector_imp} shows the cumulative importance of the vector-like engineered features: the dimensions of the projective spaces, together with the principal components of the configuration matrix, are by far the most important variables.

\begin{figure}[tbp]
    \centering
    \includegraphics[width=0.7\linewidth]{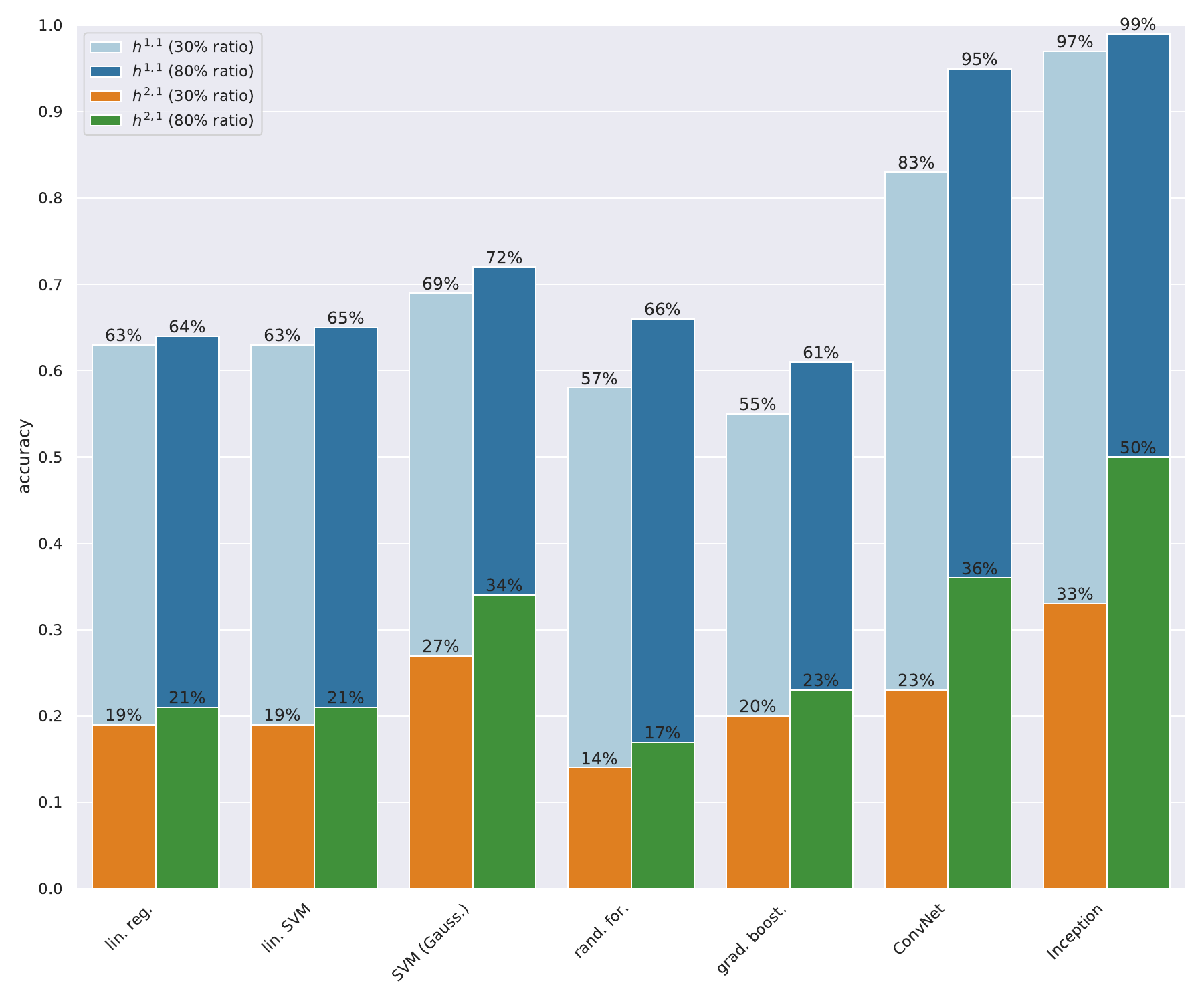}
    \caption{Comparison of performances for different ML algorithms to predict Hodge numbers in CICY 3-folds~\cite{erbin2021machine}.}
    \label{fig:cicy3-non-nn-algo}
\end{figure}

In conclusion, an accurate data analysis may recover relevant information hidden in the raw input.
While this analysis suggests taking into consideration the information encoded separately in the rows and columns of the configuration matrix, the engineered features were not found to improve the results with neural networks~\cite{erbin2021machine}.
Other ML algorithms do benefit from incorporating engineered features (\Cref{fig:cicy3-non-nn-algo}), but they are all outperformed  by neural networks, on which we focus in the following sections.

\section{Neural networks for CICYs}

In this section, we review the recently proposed neural network architectures to compute CICY Hodge numbers~\cite{erbin2021inception, erbin2021machine, erbin2021deep}.
Recently, a novel interest in convolution based architectures~\cite{lecun1989backpropagation} sprouted in applications of deep learning techniques to physics~\cite{ruehle2020data}.
Such architectures leverage the ability to approximate complex, non-linear functions with the advantages typical of computer vision and object recognition tasks.
These architectures are capable of exploring recurring patterns, through different learnable \emph{filters}, and to extract meaningful information.
They show interesting properties, such as translational equivariance~\cite{zhang1988shift, mouton2021stride}, which may, in some cases, be beneficial to pick up specific features.
Combinations of different convolution operations have shown promising results, as they enable exploration of the input at different scales and shapes~\cite{szegedy2015going}.

\subsection{The Inception Module}

\begin{figure}[tbp]
    \centering
    \includegraphics[width=0.7\linewidth]{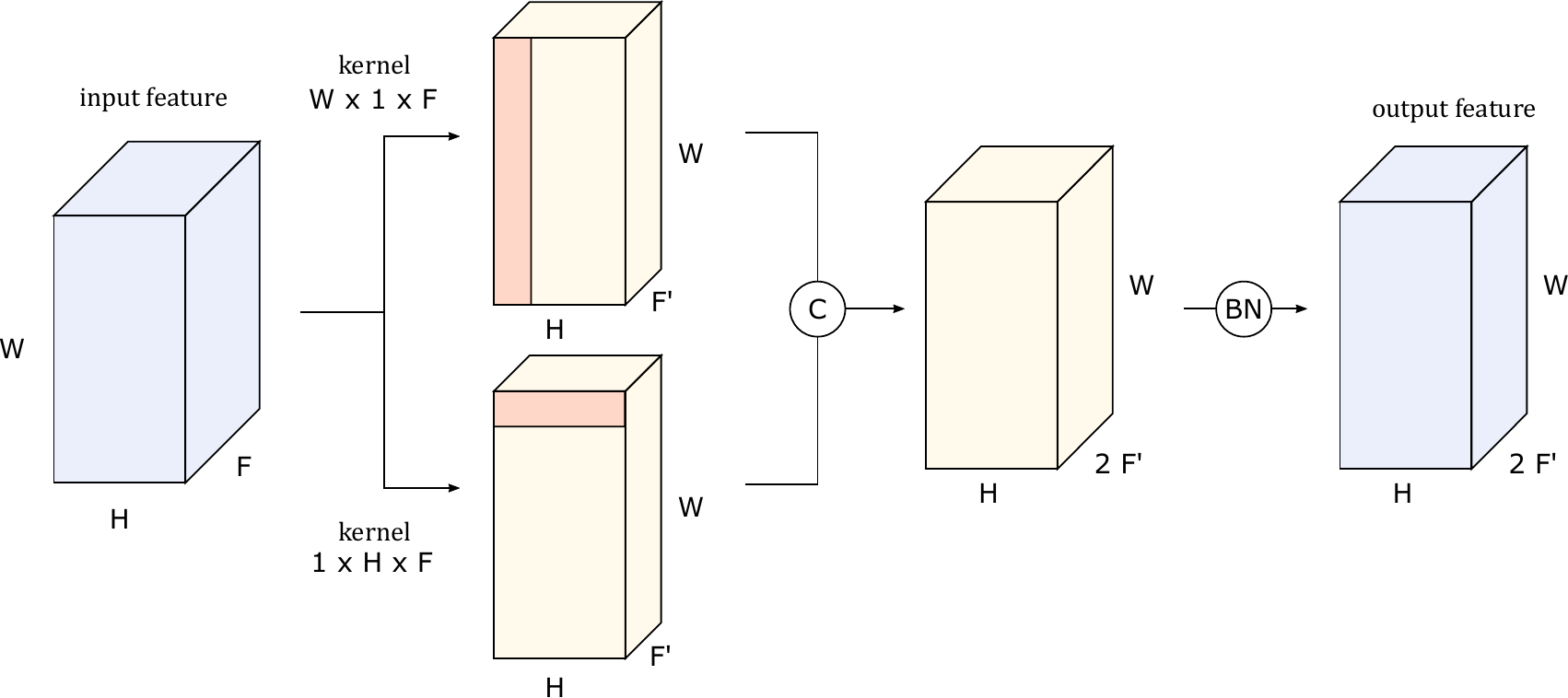}
    \caption{%
        The Inception module.
        Concatenation ($\mathrm{C}$) and batch normalisation ($\mathrm{BN}$) operations are indicated in the diagram.
    }
    \label{fig:inception}
\end{figure}

Inspired by GoogleNet's~\cite{szegedy2015going} success, the ability to combine different convolutions to compute topological quantities has been explored for the first time in~\cite{erbin2021inception, erbin2021deep}.
The \emph{Inception} module combines different kernel shapes to explore recurring patterns in the input, which corresponds here to the configuration matrix~\eqref{eq:config-matrix}.

In this case, the coefficients of a given row are associated with the degrees of the coordinates of a single projective space appearing in each equation, and correspondingly the coefficients of a given column are associated with the degrees of the coordinates of all projective spaces appearing in a single equation.
This suggests that the optimal choice is using both maximal 1-dimensional kernel in parallel~\cite{erbin2021inception, erbin2021machine}.
This choice may be motivated by relating the expression of Hodge numbers to the ambient space cohomology~\cite{erbin2021deep}: in the Koszul resolution, the projective spaces appear symmetrically in the ambient space.
Moreover, the resolution contains antisymmetric products of the spaces describing the hypersurface equations.
In both cases, the projective spaces and hypersurface equations appear in an equivalent way, such that it makes sense to choose kernels which respect this property.
\Cref{fig:inception} shows the module used in~\cite{erbin2021machine, erbin2021inception, erbin2021deep} and in the current review.
The ablation study performed in~\cite{erbin2021inception} displayed in a striking manner how both the use of Inception modules and $1d$ kernels together are necessary to reach the highest accuracy.

Notice that the two concurrent convolutions are concatenated and followed by a batch normalisation operation~\cite{ioffe2015batch}, which helps to contain the values of the activations (rectified linear units~\cite{glorot2011deep}), using the running statistics of the training set.

\subsection{The Inception Network}

In~\cite{erbin2021machine, erbin2021inception}, we designed a neural network made by a succession of Inception modules to compute the Hodge numbers of the CICY 3-folds.
With respect to previous attempts~\cite{He:2017:MachinelearningStringLandscape, Bull:2018:MachineLearningCICY}, the convolutional network profits from the parameter sharing, characteristic of the kernel operations, and it needs only few hidden layers to reach high accuracy.
In the setup of 3-folds, we used two separate networks for the predictions of \h{1}{1} and \h{2}{1}.

Though longer training time is needed for the operations, convolutions use less variables to compute the outputs.
Namely, the Inception networks in~\cite{erbin2021machine, erbin2021inception} employ \num{2.5e5} parameters for the prediction of \h{1}{1} and \num{1.1e6} for \h{2}{1}, to be compared to \num{1.6e6} for the fully connected network in~\cite{Bull:2018:MachineLearningCICY}.
The architectures allowed us to reach \SI{99}{\percent} in accuracy for the prediction of \h{1}{1} and \SI{50}{\percent} for \h{2}{1}~\cite{erbin2021machine, erbin2021inception}.

The natural following step is to compute both Hodge numbers at the same time, using the natural generalisation of the Inception network to a multi-task architecture with two output units.
However, this network is not immediately suitable for multi-output inference, and accuracy remains higher when computing a single Hodge numbers at a time.
This motivates us to test also for the 3-folds. the architecture introduced in~\cite{erbin2021deep} for the 4-folds, and which we describe in the next section.

Nonetheless, we provide a baseline of the Inception network for CICY 4-folds, too.
Basic hyperparameter optimisation led to an architecture with three hidden layers and 32, 64 and 32 filters in the channel dimension, and \num{0.2} as dropout rate.
Larger architectures do not dramatically improve the accuracy on the predictions of the Hodge numbers, thus we choose the smaller architecture as reference.
The total number of parameters is thus \num{3.8e5}.
We use \num{0.05}, \num{0.3}, \num{0.25} and \num{0.35} as loss weights for \h{1}{1}, \h{2}{1}, \h{3}{1} and \h{2}{2}, respectively.
Training lasted \num{1500} epochs on a single NVIDIA V100 GPU, with a starting learning rate of \num{e-3} and the Adam~\cite{kingma2014adam} optimiser, and a mini-batch size of 64 configuration matrices.
The learning rate is reduced by a factor \num{0.3} after \num{450} epochs without improvement in the total validation loss.

\subsection{CICYMiner Network}

\begin{figure*}
    \centering
    \includegraphics[width=0.8\linewidth]{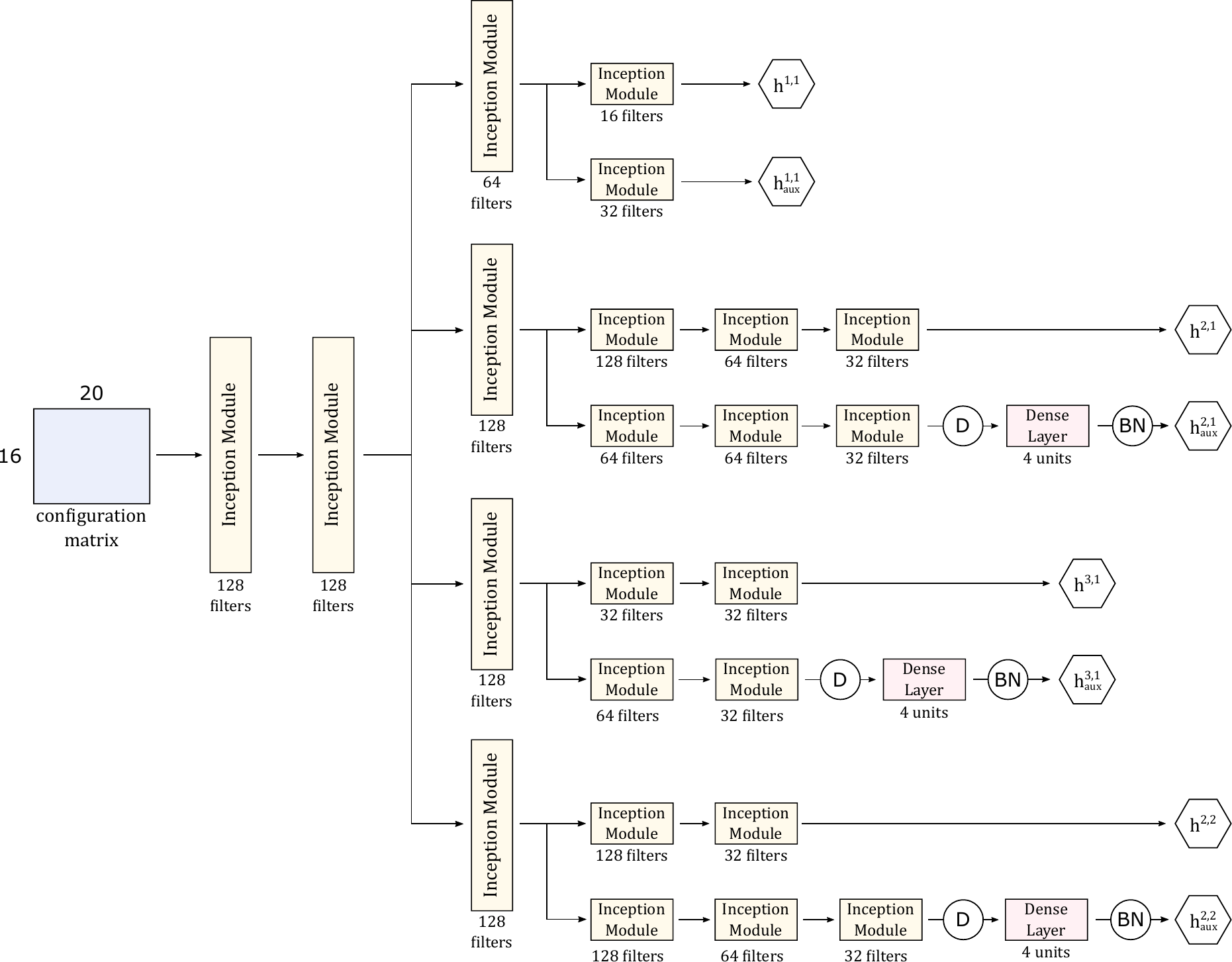}
    \caption{%
        CICYMiner architecture.
        Dropout ($\mathrm{D}$) and batch normalisation ($\mathrm{BN}$) operations are indicated.
    }
    \label{fig:cicyminer}
\end{figure*}

CICYMiner is an architecture introduced in~\cite{erbin2021deep} to compute Hodge numbers of CICY 4-folds.
In~\Cref{fig:cicyminer}, we show the original architecture used for simultaneous predictions of the four non-trivial Hodge numbers.
The architecture enables multi-task learning via hard parameter sharing~\cite{ruder2017overview}: the output predictions are computed from a common shared representation of the input through differentiated branches.
This approach has proven efficient in increasing the learning power of the network, and in reducing the risk of overfitting~\cite{caruana1997multitask, baxter1997bayesian}.
The intermediate layers of the network replicate the same structure by introducing an auxiliary branch, which replicates the output predictions.
In turn, this helps the stability of the learning process.
It also enables the \emph{mining}~\cite{benzine2021deep} of richer and diverse features from a shared feature map, that is the computations of different characteristics at different levels and scales, in order to enrich the information extracted by each branch.
CICYMiner is thus capable of leveraging the extraction of a larger number of features with the advantages of multi-task learning.
Specifically, handling outliers is simpler in a multi-task architecture, as their presence for a specific output may help the prediction of a different task.
We thus choose to keep the outliers in the training and validation sets for CICYMiner and, compared to~\cite{erbin2021inception,erbin2021machine}, we find that they don't decrease the performance.
The preprocessing of Hodge numbers in~\cite{erbin2021inception} becomes less impacting with this specific architecture.
A simple dropout rate of \num{0.2} has been added before the dense layers to avoid a strong architecture dependence, and to introduce a degree of randomness during training.
The network in \Cref{fig:cicyminer} contains approximately \num{e7} parameters.

Since the prediction of the Hodge numbers is a regression task, the output of CICYMiner are positive floating point numbers (ensured by adding a rectified linear unit to each output layer).
As they need to be compared to integers, predictions are rounded to the closest integer before computing the accuracy of the network.
The learning process is complemented by the choice of the \emph{Huber} loss function~\cite{huber1992robust}:
\begin{equation}
    \small
    \mathcal{H}^{\{ k \}}_{\delta}( x )
    =
    \begin{cases}
        \displaystyle
        \frac{1}{2}
        \sum\limits_{n = 1}^k
        \sum\limits_{i = 1}^{N_k}\,
        \omega_n
        \big( x^{(i)} \big)^2,
        &
        \left| x^{(i)} \right|
        \le \delta
        \\
        \displaystyle
        \delta
        \sum\limits_{n = 1}^k
        \sum\limits_{i = 1}^{N_k}\,
        \omega_n
        \left(
            \big| x^{(i)} \big|
            -
            \frac{\delta}{2}
        \right),
        &
        \left| x^{(i)} \right|
        > \delta
    \end{cases}
    .
\end{equation}
The choice of the learning objective is, again, dictated by the handling of outliers, through the introduction of a sparsity filter for largely diverging predictions.
Robustness is implemented as an interpolation between the quadratic and linear behaviour of the function.
This solution has already been adopted in classification tasks~\cite{pan2009survey}, where different combinations of $\ell_1$ norm, $\ell_2$ norm and Frobenius norm are used for robustness.
The parameter $\delta$ is an additional hyperparameter of the model.
Regression metrics such as the mean squared error (MSE) and the mean absolute error (MAE) can also be used to get more indications on the learning process: in particular, they can show whether the network is actually capable of learning the discreteness of the Hodge numbers.

As mentioned in the previous subsection, we will test the same architecture in~\Cref{fig:cicyminer} for predicting the Hodge numbers of the 3-folds, which is a new investigation.
This is achieved by removing the legs for $\h{2}{2}$ and $\h{3}{1}$.
In total, we have \num{3.3e6} parameters for the 3-folds, which is comparable to the two Inception networks in~\cite{erbin2021machine} combined (\num{2.5e5} parameters for \h{1}{1} and \num{1.1e6} for \h{2}{1}).
Basic optimisation of the hyperparameters enabled a reduction of the parameters by a factor \num{3} without strongly impacting the accuracy.
For simplicity, we will keep the same CICYMiner architecture of~\cite{erbin2021deep}.

We preprocess the input data by simply rescaling the entries of the configuration matrices in the training set in the $[0, 1]$ range.
Notice that, even though the ranges of definition strongly differ amongst the Hodge numbers (see~\Cref{fig:cicy4_hodge}), there is no need to rescale the labels, as the deep structure of CICYMiner is capable of handling such differences.

Training has been performed on a single NVIDIA V100 GPU over a fixed amount of \num{300} epochs for CICY 4-folds, while we use \num{1500} epochs in the 3-folds case, due to the limited cluster time.
We use the Adam~\cite{kingma2014adam} optimiser with an initial learning rate of \num{e-3} and a mini-batch size of \num{64} (bs-64) configuration matrices.
The learning rate is reduced by a factor \num{0.3} after \num{75} epochs without improvements in the total loss of the validation set.
Due to limited time, hyperparameter optimisation is performed on the 4-folds dataset using a grid search over a reasonable portion of the hyperparameter space.
The implementation uses $\delta = 1.5$ and loss weights of \num{0.05}, \num{0.3}, \num{0.25} and \num{0.35} for \h{1}{1}, \h{2}{1}, \h{3}{1} and \h{2}{2}, respectively, in the 4-folds case.
In order to maintain the same ratio, we use \num{0.17} and \num{0.83} for \h{1}{1} and \h{2}{1} for the CICY 3-folds.

\section{Applications}
\label{sec:applications}

We provide three applications of the CICYMiner network.
In \Cref{sec:training-any}, we reproduce the predictions for the Hodge numbers of the CICY 4-folds from~\cite{erbin2021deep} and generalize to the 3-folds.
Next, in \Cref{sec:training-low,sec:training-high}, we study how CICYMiner can extrapolate its predictions after training only with low/high Hodge numbers.
Except for the 4-fold predictions in \Cref{sec:training-any}, the results in \Cref{sec:applications} are new.

\subsection{Learning Hodge numbers}
\label{sec:training-any}

As a first application, we are interested in predicting the Hodge numbers, using the configuration matrix as input of the deep architecture.
For both 3-folds and 4-folds we select the training set via a stratifies approach on \h{2}{1} in order to preserve the distribution of the samples: the CICY 4-folds dataset is, in fact, highly peculiar, since \h{2}{1} vanishes for \SI{70}{\percent} of the configuration matrices.
Though the effect is much less salient for the 3-folds, we adopt the same strategy for comparison.
The validation set is chosen totally at random, using \SI{10}{\percent} of the samples.
The remaining samples are included in the test set.

\begin{figure}[tbp]
    \centering
    \includegraphics[width=0.7\linewidth]{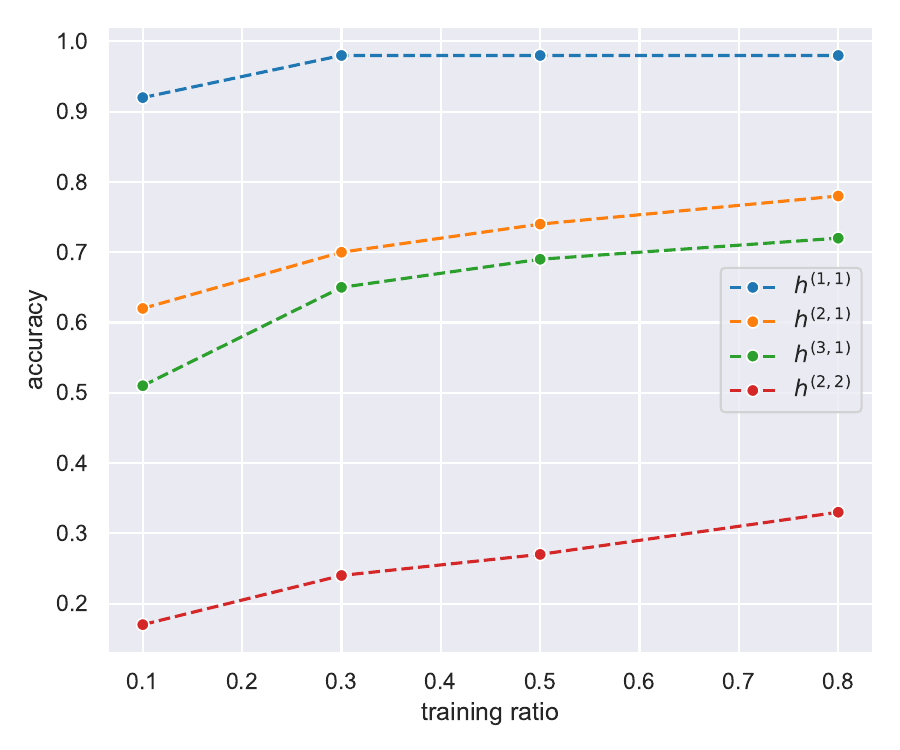}
    \caption{Learning curve of Inception network on the CICY 4-folds dataset.}
    \label{fig:cicy4_lc_inception}
\end{figure}

\begin{figure}[tbp]
    \centering
    \includegraphics[width=0.7\linewidth]{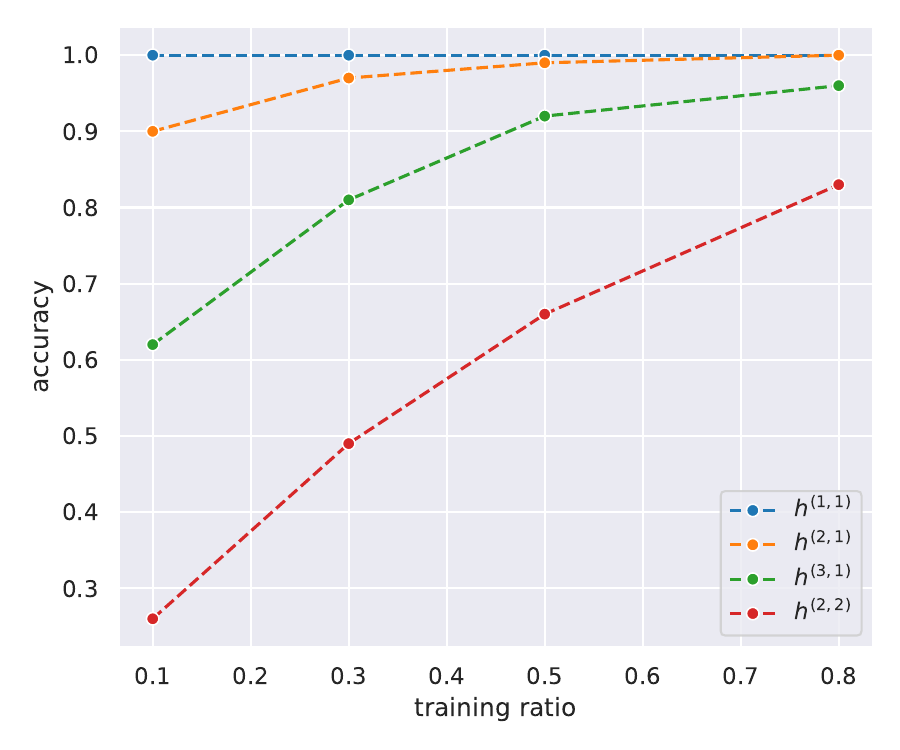}
    \caption{Learning curve of CICYMiner on the CICY 4-folds dataset.}
    \label{fig:cicy4_lc}
\end{figure}

\begin{table}[tbp]
    \centering
    \begin{tabular}{@{}lcccc@{}}
    \toprule
               & \h{1}{1}      & \h{2}{1}      & \h{3}{1}      & \h{2}{2}      \\ \midrule
    +att       & \textbf{1.00} & 0.99          & \textbf{0.96} & 0.81          \\
    MSE loss   & \textbf{1.00} & 0.97          & 0.92          & 0.50          \\
    no aux     & \textbf{1.00} & 0.84          & 0.92          & 0.72          \\
    bs-256     & \textbf{1.00} & 0.99          & 0.94          & 0.65          \\
    layer norm & \textbf{1.00} & 0.99          & 0.92          & 0.66          \\ \midrule
    \textbf{CICYMiner}      & \textbf{1.00} & \textbf{1.00} & \textbf{0.96} & \textbf{0.83} \\
    \small{~~MSE (\num{e-4})} & \num{1.3}  & \num{98}  & \num{560}  & \num{6800}            \\
    \small{~~MAE (\num{e-3})} & \num{7.8}  & \num{19}  & \num{130}  & \num{360}             \\ \midrule
    \textbf{Inception}      & 0.98 & 0.78 & 0.72 & 0.33 \\ \bottomrule
    \end{tabular}
    \caption{%
        Ablation study at \SI{80}{\percent} training ratio for the CICY 4-folds.
        The metric displayed is the accuracy.
    }
    \label{tab:cicy4_acc_table}
\end{table}

In~\Cref{fig:cicy4_lc_inception}, we present the learning curve of the Inception network on the 4-folds dataset, used as baseline for comparisons with the CICYMiner architecture.
In~\Cref{fig:cicy4_lc}, we show the learning curve of CICYMiner on the 4-folds dataset.
As visible, with CICYMiner, \h{1}{1} reaches perfect accuracy with just \SI{10}{\percent} of the data.
This behaviour is partially due to the presence of several configuration matrices in the favourable representation (\SI{54.5}{\percent}), for which \h{1}{1} is just the number of projective ambient space factors, $\h{1}{1} = r$.

\begin{figure}[tbp]
    \centering
    \includegraphics[width=0.7\linewidth]{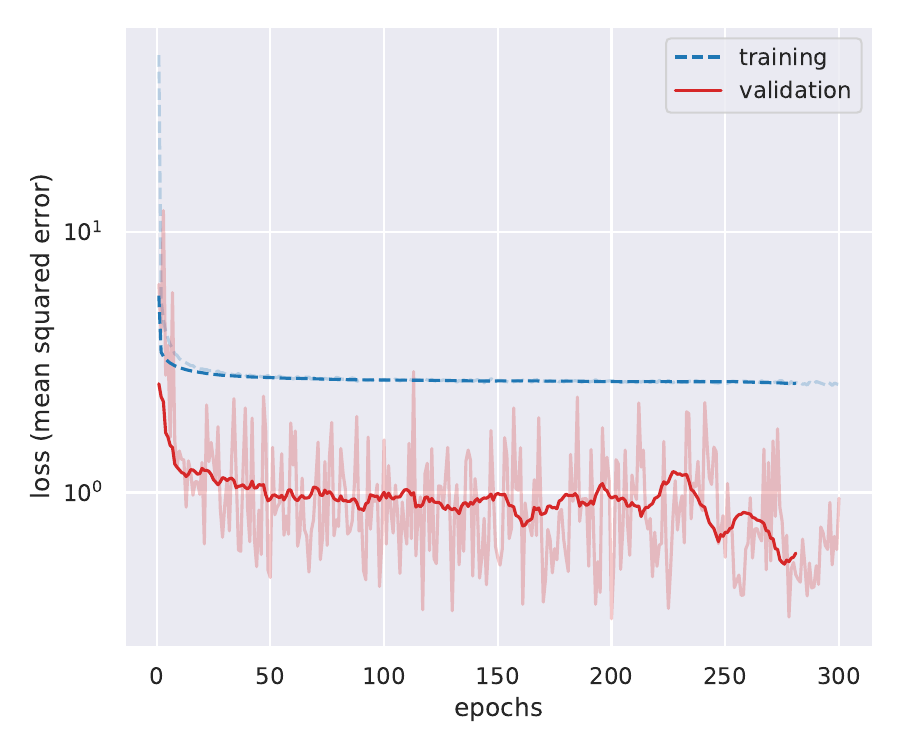}
    \caption{Loss function of CICYMiner on the CICY 4-folds dataset at \SI{80}{\percent} training ratio.}
    \label{fig:cicy4_loss}
\end{figure}

In~\Cref{tab:cicy4_acc_table}, we briefly summarise the ablation study performed for the CICY 4-folds at \SI{80}{\percent} of training ratio.
The CICYMiner network is also capable of learning accurately the discreteness of the Hodge numbers: for most of them, the regression metrics show values which can confidently indicate integer numbers (MAE $\ll 0.50$ and MSE $\ll 0.25$).
As shown in~\Cref{fig:cicy4_loss}, the network is still underfitting the training set: longer training or different choices of the learning rate may be needed to investigate this aspect.

\begin{figure*}
    \centering
    \begin{subfigure}[b]{0.7\linewidth}
    \centering
    \includegraphics[width=\linewidth]{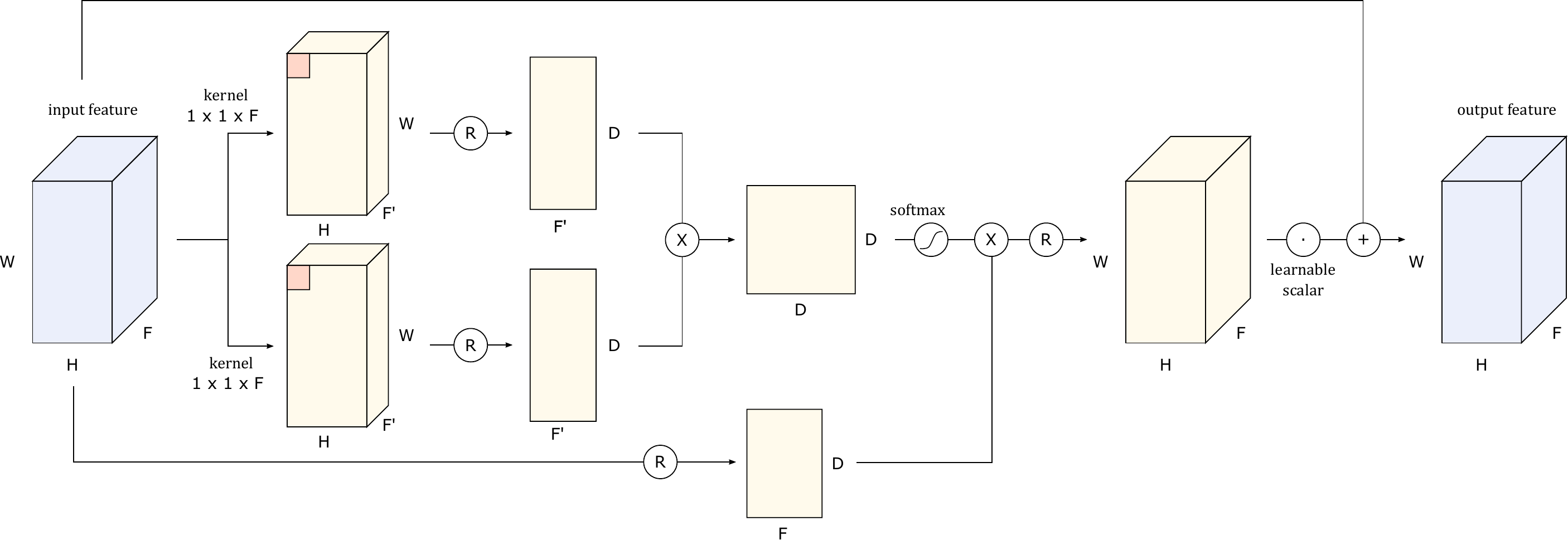}
    \caption{Spatial Attention Module}
    \end{subfigure}
    \\
    \begin{subfigure}[b]{0.7\linewidth}
    \centering
    \includegraphics[width=\linewidth]{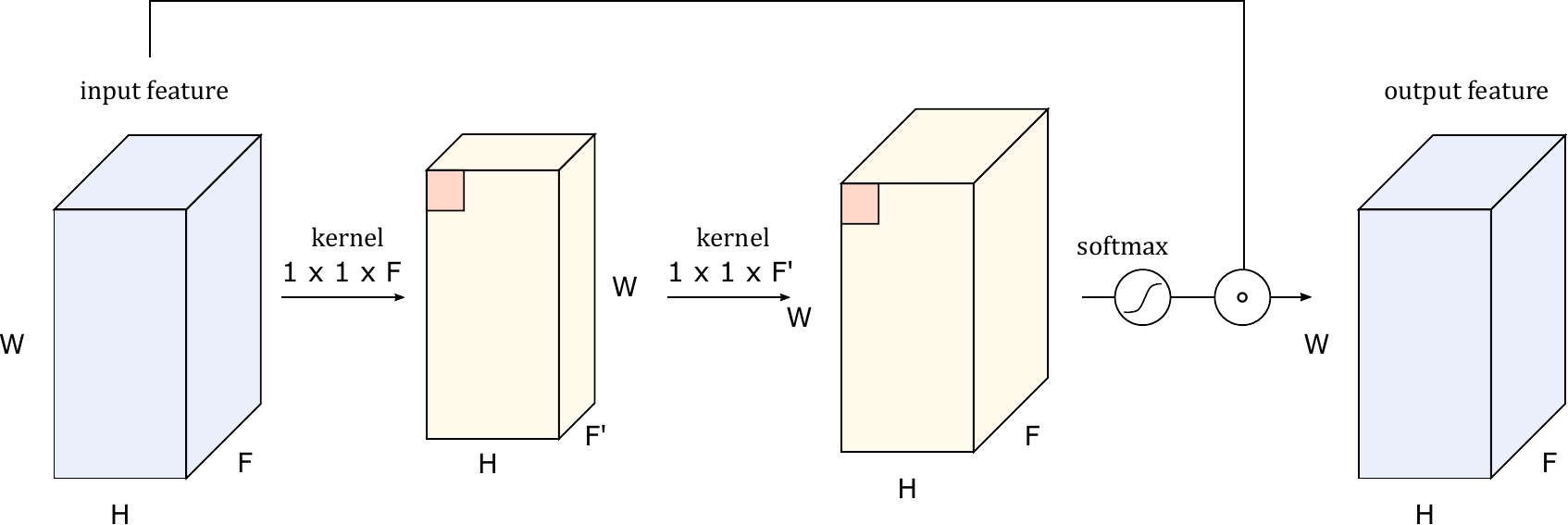}
    \caption{Channel Attention Module}
    \end{subfigure}
    \caption{%
        Attention modules used in the ablation study.
        Here, $\times$ indicates a matrix product along the appropriate axes, while $\circ$ is the Hadamard product.
        Reshape operations ($\mathrm{R}$) are also indicated.
    }
    \label{fig:attention}
\end{figure*}

The ablation study was performed by exploring different architectures, such as the addition of attention modules used in~\cite{benzine2021deep} after each Inception module in the auxiliary branch, the use of the traditional $\ell_2$ loss, the absence of the auxiliary branches, an increased mini-batch size (bs), and a different normalisation strategy based on layer normalisation~\cite{ba2016layer}.
The principal advantage of the introduction of an attention mechanism, made by the composition of a channel attention module (CHAM) and spatial attention module (SAM) shown in~\Cref{fig:attention}, is the faster convergence of the loss function~\cite{erbin2021deep}, but it does not lead to better results.
On the contrary, the $\ell_2$ loss shows a strong drop in accuracy in the presence of many outliers, as happens for instance for \h{2}{2}.
The ablation study also shows the function of the auxiliary branches, as their introduction is capable of mining richer information, which in turn is beneficial to \h{2}{1}, as it boosts the prediction ability in the case of the ``imbalanced class'' (the term is used by extension, as this is not a classification task).
The normalisation strategies are also relevant, as the use of a non-trivial mini-batch size is capable of retaining partial information on the physics contained in the dataset.
However, larger sizes can lead to a decrease in performance, as the large heterogeneity of the data may damage the stability of the predictions.

\begin{figure}[tbp]
    \centering
    \includegraphics[width=0.7\linewidth]{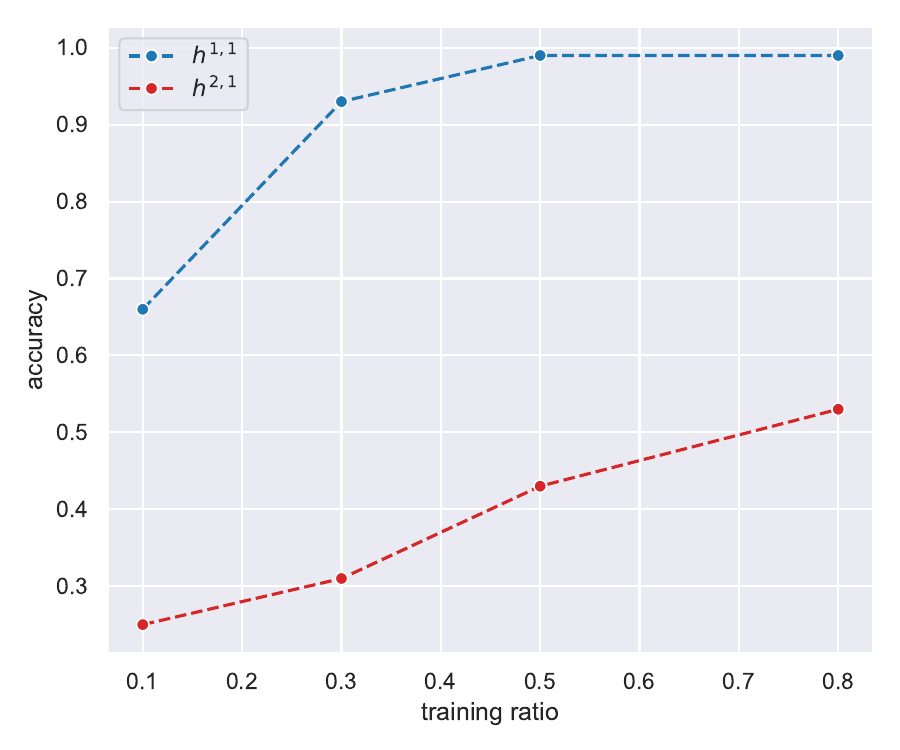}
    \caption{Learning curve of CICYMiner on the CICY 3-folds dataset.}
    \label{fig:cicy3_lc}
\end{figure}

The same experiment can be run on CICY 3-folds.
Since the dataset is vastly smaller than the 4-folds dataset, we opt for a longer training (\num{1500} epochs) to test the convergence.
In~\Cref{fig:cicy3_lc} we show the learning curve associated to the training procedure: the architecture quickly reaches very high accuracy for \h{1}{1}, while \h{2}{1} remains quite difficult to reproduce.

\begin{table}[tbp]
    \centering
    \begin{tabular}{@{}lcc@{}}
    \toprule
               & \h{1}{1}      & \h{2}{1}     \\ \midrule
    +att       & \textbf{0.99} & 0.47          \\
    MSE loss   & 0.98          & 0.44          \\
    no aux     & 0.98          & 0.37          \\
    bs-8       & 0.99          & 0.52          \\
    bs-256     & 0.98          & 0.37          \\
    layer norm & 0.98          & 0.17          \\ \midrule
    \textbf{CICYMiner}      & \textbf{0.99} & \textbf{0.53}    \\
    \small{~~MSE (\num{e-2})} & \num{5.3}  & \num{100} \\
    \small{~~MAE (\num{e-2})} & \num{4.8}  & \num{70} \\  \midrule
    \textbf{Inception}      & \textbf{0.99} & 0.50    \\ \bottomrule
    \end{tabular}
    \caption{%
        Ablation study at \SI{80}{\percent} training ratio for the CICY 3-folds.
        The metric displayed is the accuracy.
    }
    \label{tab:cicy3_acc_table}
\end{table}

\Cref{tab:cicy3_acc_table} shows the results of the ablation study performed on the CICY 3-folds.
In this scenario, the reduced size of the training set (the full CICY 3-folds database is a factor \num{e2} smaller than the CICY 4-folds dataset) impacts negatively on the ability to learn the Hodge numbers.
Though not fully optimised for the task, results are comparable with previous attempts using Inception-based architectures~\cite{erbin2021inception, erbin2021machine}.

\begin{figure}[tbp]
    \centering
    \includegraphics[width=0.7\linewidth, trim={0 0 0 0.35in}, clip]{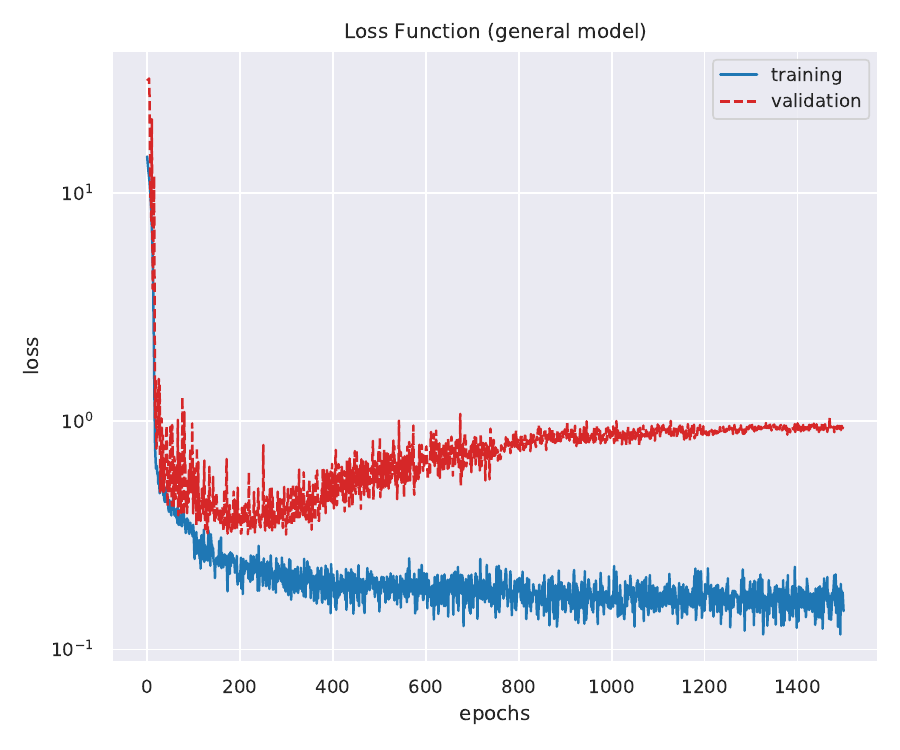}
    \caption{Loss function of CICYMiner on the CICY 3-folds dataset at \SI{80}{\percent} training ratio.}
    \label{fig:cicy3_loss}
\end{figure}

The regression metrics MAE and MSE show that only \h{1}{1} has been effectively learnt as an integer number, but, the discreteness of \h{2}{1} is instead quite difficult to recover using a small dataset.
Moreover, the use of a robust training loss makes up for the lack of hyperparameter optimisation: using just a $\ell_2$ loss shows a strong decrease in the accuracy of \h{2}{1}.
We also notice that, in this case, the distribution of the configuration matrices seems to be non-homogeneous: larger mini-batch sizes seem to spoil the ability to provide good predictions for \h{2}{1}, while the layer normalisation strategy leads to worst results.
In this case, given the smaller dataset, we also provide the ablation study with a small batch size, which shows results comparable with CICYMiner.
The loss function shown in~\Cref{fig:cicy3_loss} also shows a difficult training procedure: the loss on the validation set is unstable, due to the reduced number of configuration matrices.
It also shows an increase towards the end of the training procedure.

Overall, the CICYMiner architecture shows good results in the presence of a large number of configuration matrices in the training set.
It also retains generalisation capabilities, as no additional hyperparameter optimisation seems to be needed to switch from the 4-folds dataset to the 3-folds dataset.

\subsection{Training at low Hodge numbers}
\label{sec:training-low}

Similar to the previous learning task, we focus on the CICY 4-folds dataset.
Following~\cite{Bull:2019:GettingCICYHigh}, we test the ability to predict the entire range of Hodge numbers when using only a limited range for training.
That is, we restrict the variability of the Hodge numbers, specifically \h{1}{1}, by imposing an upper bound, in order to train only on small values.
Then, we test how well the network performs in predicting Hodge numbers outside the training range.
We reuse the same hyperparameters as before, but train CICYMiner on the reduced training set.
Results are summarised in~\Cref{tab:cicy4_train_low}.
Here, as in the following, the ratio of the training data refers to the full training data used in the previous section.

\begin{table}[tbp]
    \centering
    \begin{tabular}{@{}lccccc@{}}
    \toprule
                      & ratio              & \h{1}{1}   & \h{2}{1}   & \h{3}{1}   & \h{2}{2}   \\ \midrule
    $\h{1}{1} \le 5$  & \SI{2}{\percent}   & 0.06       & 0.70       & 0.07       & 0.02       \\
    $\h{1}{1} \le 8$  & \SI{26}{\percent}  & 0.53       & 0.83       & 0.27       & 0.11       \\
    $\h{1}{1} \le 10$ & \SI{59}{\percent}  & 0.93       & 0.95       & 0.62       & 0.40       \\ \bottomrule
    \end{tabular}
    \caption{Accuracy of CICYMiner for CICY 4-folds on different ranges of values of \h{1}{1}.}
    \label{tab:cicy4_train_low}
\end{table}

We arbitrarily choose \num{5}, \num{8}, \num{10} as upper bounds of \h{1}{1} for training.
As expected, the larger the reduction in training samples, the lower the accuracy achieved by CICYMiner.
Notice that in the $\h{1}{1} \le 5$ case, the high residual accuracy on \h{2}{1} reflects the large amount of vanishing labels, as the network is mostly predicting $0$ for the Hodge number.
Notice also that, in these cases, the range of \h{2}{2} is mostly unaffected by the chosen intervals of \h{1}{1}: its range of variation is constantly $[204, 1752]$, which shows a more complicated dependence of \h{2}{2} on \h{1}{1} with respect to other Hodge numbers.
For the other Hodge numbers, the choice of the intervals of \h{1}{1} mostly impacts the lower bound of \h{3}{1} which varies as $[35, 426]$, $[32, 426]$ and $[30, 426]$ in the three cases of~\Cref{tab:cicy4_train_low}.
The upper bound of \h{2}{1} is also slightly concerned, as it is limited to $[0, 30]$.

In general, this experiment shows complicated relations between the Hodge numbers, particularly when referred to \h{1}{1}.
Nonetheless, good results can be obtained for some of them even when using smaller datasets.
For instance, \h{1}{1} and \h{2}{1} gain quite good accuracy rapidly, even with \SI{60}{\percent} of the training ratio, dictated by a cut-off on \h{1}{1}.

\begin{table}[tbp]
    \centering
    \begin{tabular}{@{}lccc@{}}
    \toprule
                      & ratio              & \h{1}{1}   & \h{2}{1}   \\ \midrule
    $\h{1}{1} \le 3$  & \SI{3}{\percent}   & 0.03       & 0.00       \\
    $\h{1}{1} \le 5$  & \SI{20}{\percent}  & 0.04       & 0.18       \\
    $\h{1}{1} \le 7$  & \SI{54}{\percent}  & 0.50       & 0.34       \\
    $\h{1}{1} \le 8$  & \SI{70}{\percent}  & 0.81       & 0.42       \\ \bottomrule
    \end{tabular}
    \caption{Accuracy of CICYMiner for CICY 3-folds on different ranges of values of \h{1}{1}.}
    \label{tab:cicy3_train_low}
\end{table}

A similar approach can be used for CICY 3-folds.
In~\Cref{tab:cicy3_train_low} we show the results of the training at low Hodge numbers in the case of three complex dimensions.
We see that the particular choice of the training set strongly impacts the ability of the network to reach high accuracy for both Hodge numbers: this is due to the hard cut-off imposed (compare with~\Cref{fig:cicy3_hodge}, for instance), which does not allow the network to correctly predict the out-of-distribution samples.
Notice that, in general, the range of \h{2}{1} slightly changes under the different choices of \h{1}{1}: for an upper bound on the first Hodge number of \num{3}, \num{5}, \num{7} and \num{8}, the range of \h{2}{1} is modified into $[27, 101]$, $[25, 101]$, $[23, 101]$ and $[22, 101]$, respectively.
With respect to~\cite{Bull:2019:GettingCICYHigh}, the accuracy reached by CICYMiner is in any case higher and increases consistently when the size of the training set is increased.
Curiously, at very low training ratio, the accuracy on the prediction of the Hodge numbers can be greatly increased by flooring the predictions (rather than approximating to the nearest integer).
For example, when considering $\h{1}{1} \le 5$, the accuracy for \h{1}{1} rises to \SI{19}{\percent}, contrary to what shown in~\Cref{tab:cicy3_train_low} (the accuracy of \h{2}{1} drops to \SI{16}{\percent}, in this case).
This may be an indication that the network is over-estimating the values of the Hodge numbers when smaller training sets are used.

\subsection{Training at high Hodge numbers}
\label{sec:training-high}

The same experiment can be run in another interesting case, imposing a lower bound on \h{2}{2}.
Since the latter is the Hodge number with the largest number of outliers and the largest interval of variation, controlling its range may lead to some interesting relations.
As previously, no hyperparameter optimisation is run in this case.
Results are shown in~\Cref{tab:cicy4_train_high}, where the ratio of the data refers to the size of the training set used in~\Cref{sec:training-any}.

\begin{table}[tbp]
    \centering
    \begin{tabular}{@{}lccccc@{}}
    \toprule
                      & ratio               & \h{1}{1}    & \h{2}{1}   & \h{3}{1}   & \h{2}{2} \\ \midrule
    $\h{2}{2} > 225$  & \SI{50}{\percent}   & 1.00       & 0.88       & 0.39       & 0.04      \\
    $\h{2}{2} > 250$  & \SI{27}{\percent}   & 0.98       & 0.80       & 0.09       & 0.03      \\
    $\h{2}{2} > 300$  & \SI{8}{\percent}    & 0.95       & 0.72       & 0.05       & 0.03      \\
    $\h{2}{2} > 400$  & \SI{1.6}{\percent}  & 0.76       & 0.70       & 0.03       & 0.0       \\ \bottomrule
    \end{tabular}
    \caption{Accuracy of CICYMiner for CICY 4-folds on different ranges of values of \h{2}{2}.}
    \label{tab:cicy4_train_high}
\end{table}

The lower bounds on \h{2}{2} were arbitrarily chosen as \num{225}, and \num{400}.
This choice generally strongly impacts the lower bound on \h{3}{1}, which becomes $28$, $79$, $158$ and $211$ respectively.
Moreover, the lower bound of \h{2}{1} is reduced from \num{33} to \num{0}, as well as the upper bound on \h{1}{1} from \num{24} to \num{5}.
Good accuracy on \h{2}{1} are, again, due to the prevalence of vanishing labels.

As in the previous case, good results on \h{1}{1} and \h{2}{1} can be recovered starting from approximately \SI{50}{\percent} of the training samples.
The same complicated relations between the Hodge numbers and the configuration matrices are, once again, evident.

\section{Conclusions}

\begin{figure}[tbp]
    \centering
    \includegraphics[width=0.5\linewidth]{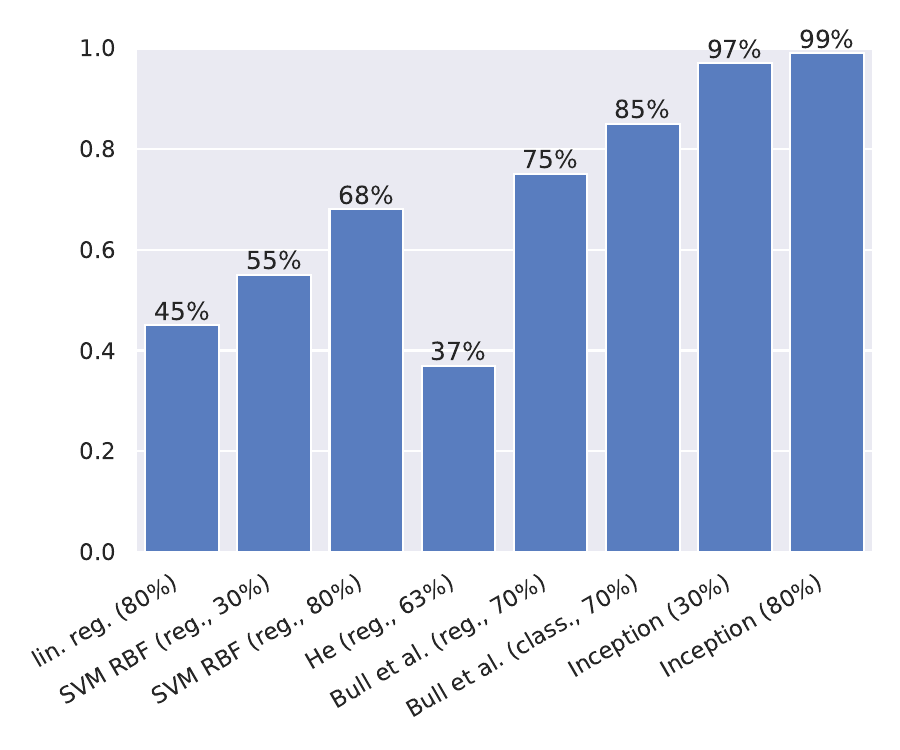}
    \caption{Comparison of accuracy for $h^{1,1}$ for CICY 3-folds using different methods~\cite{erbin2021inception, erbin2021machine}. “He” refers to the neural network in~\cite{He:2017:MachinelearningStringLandscape} (the accuracy is the one quoted in~\cite{Bull:2018:MachineLearningCICY}), and “Bull et al.” to the one in~\cite{Bull:2018:MachineLearningCICY}.}
    \label{fig:cicy3_h11_soa}
\end{figure}

In our series of works~\cite{erbin2021inception, erbin2021machine, erbin2021deep}, we have established the state of the art for the computations of Hodge numbers of CICY using machine learning.
The Hodge number which has been the most studied is $\h{1}{1}$, and we illustrate the improvement in predicting it over time and using different methods in \Cref{fig:cicy3_h11_soa}.

Computing accurately \h{2}{1} for the 3-folds is still an open problem.
While the CICYMiner architecture performs slightly better than the pure Inception model, the drawback is a much slower training time and many more parameters.
Moreover, while results are more precise for the 4-folds, it would be desirable to increase the accuracy of \h{2}{2}.
As suggested in~\cite{erbin2021inception, erbin2021machine}, a possibility is to represent the configuration matrix by a graph~\cite{hubsch1992calabi, Krippendorf:2020:DetectingSymmetriesNeural}.

Now that most Hodge numbers can be well computed using deep learning, it would be interesting to extract additional analytic information.
One possibility is to use symbolic regression for neural networks as developed in~\cite{Cranmer:2019:LearningSymbolicPhysics, Cranmer:2020:DiscoveringSymbolicModels} (noting that the authors point out that graph neural networks are much better in this context as they can encode substructure in a finer way), using algorithms such as \texttt{eureqa}~\cite{Schmidt:2009:DistillingFreeFormNatural} or \texttt{AIFeynman}~\cite{Udrescu:2020:AIFeynmanPhysicsInspired, Udrescu:2020:AIFeynman20}.
In particular, \texttt{eureqa} can learn piecewise functions~\cite{Cranmer:2020:DiscoveringSymbolicModels}, which is particularly relevant given the structure of the analytic expressions found for line bundle cohomologies~\cite{Constantin:2019:FormulaeLineBundle, Larfors:2019:LineBundleCohomologies, Klaewer:2019:MachineLearningLine, Brodie:2020:IndexFormulaeLine, Brodie:2020:MachineLearningLine}.

\section*{Acknowledgments}

We are grateful to Robin Schneider for past collaborations and many discussions on this topic.
This project has received funding from the European Union's Horizon 2020 research and innovation program under the Marie Skłodowska-Curie grant agreement No 891169.
This work is supported by the National Science Foundation under Cooperative Agreement PHY-2019786 (The NSF AI Institute for Artificial Intelligence and Fundamental Interactions, \url{http://iaifi.org/}).
This publication was made possible by the use of the \emph{FactoryIA} supercomputer, financially supported by the \emph{Île-De-France Regional Council}.

\printbibliography[heading=bibintoc]

\end{document}